\documentclass[10pt,aps,pra,groupedaddress,nofootinbib]{revtex4-1}
\pdfoutput=1
\usepackage{graphicx}
\usepackage{xcolor}
\usepackage[english]{babel}

\newcommand{\mm}{\,$\mu$m}
\begin{document}


\title{Pore cross-talk in colloidal filtration}


\author{Olivier Liot}
\altaffiliation[]{F\'ed\'eration FERMaT, INP Toulouse}
\email[]{olivier.liot@laas.fr}
\author{Akash Singh}%
\author{Patrice Bacchin}
\altaffiliation[]{Laboratoire de G\'enie Chimique, Universit\'e de Toulouse, CNRS, Toulouse, France}
\author{Paul Duru}
\altaffiliation[]{Institut de M\'ecanique des Fluides, IMFT, Universit\'e de Toulouse, CNRS - Toulouse, France}
\author{Jeffrey F. Morris}
\altaffiliation[]{Levich Institute and Chemical Engineering, CUNY City College of New York, USA}
\author{Pierre Joseph}
\email[]{pierre.joseph@laas.fr}
\affiliation{LAAS-CNRS, Universit\'e de Toulouse, CNRS, Toulouse, France}
%


\date{\today}

\begin{abstract}
Blockage of pores by particles is found in many processes, including filtration and oil extraction.
We present \textcolor{black}{filtration} experiments through a linear array of ten channels with one dimension which is sub-micron, through which a dilute dispersion of Brownian polystyrene spheres flows \textcolor{black}{under the action of a fixed pressure drop}.
\textcolor{black}{The growth rate of a clog formed by particles at a pore entrance} systematically increases with the number of  \textcolor{black}{already} saturated (entirely clogged) pores, indicating that there is an interaction or ``cross-talk'' between the pores. 
This observation is interpreted based on \textcolor{black}{a phenomenological  model, stating that  a diffusive redistribution of particles occurs along the membrane, from clogged to free pores.} This one-dimensional model could be extended to two-dimensional membranes.
\end{abstract}


\maketitle


\textbf{Introduction} 

A colloidal suspension flowing through a pore network often results in fouling or clogging. In industrial (oil recovery \cite{tavakkoli2015}, inkjet printing \cite{fuller2002}, filtration), biological (artery diseases \cite{rothberg2013}, detection of cells \cite{pang2015}) and natural (water infiltration in soils \cite{zhang2012}, precipitation inside rocks \cite{hilgers2002}) processes, the phenomenon of particle accumulation is involved. Recent improvements in visualization of suspended particles in model pores have led to new insight into the physical parameters at play in particle capture and clogging in pores \cite{dressaire2017}. There are several different clogging mechanisms. Size exclusion or sieving occurs when particles block a pore smaller than their diameter \textcolor{black}{\cite{sauret2014}}. If the pore size is larger than the particle, clogging can occur by two routes, either through particles forming an arch at the entrance of the pore \textcolor{black}{\cite{zuriguel2014}} or progressively adhering to walls and previously deposited particles, leading to blockage of the pore \textcolor{black}{\cite{wyss2006,dersoir2015}}. During the last decade, following an early study \cite{wyss2006} which described clogging of pores by smaller particles, a number of studies have focused on determining the pore-scale mechanisms involved in this form of pore blockage (e.g. \cite{bacchin2011,duru2015,robert_de_saint_vincent2016,dersoir2017,cejas2017,kim2017}). Other studies have proposed explanations of clogging using transition-state theory \cite{laar2016} or by relating it to jamming phenomena \cite{sendekie2016}.

Prior to the advent of pore-scale investigations, which have been greatly facilitated by microfluidic technology, numerous studies were made at a more macroscopic membrane scale, where the usual focus was on the ``filtration cake'' \textcolor{black}{\cite{brenner1961,ghidaglia1996,narayan1997,hong1997,hieke2009}}. Since a typical filtration membrane consists of a large number of closely-spaced pores, clog formation at one pore could affect its neighbours, and hence the macroscopic behavior of the membrane.  Considered in this way, there is a notable lack of information related to clog formation at the pore scale, with connection to the membrane scale by consideration of interactions between pores. In this work, we address this gap of knowledge at an intermediate scale, by considering in detail the time evolution of the clog formation process at pore scale, in a short one-dimensional (1-D) array of pores. We describe the interaction between pores as ``cross-talk''. While one recent paper \cite{laar2016} shows that a filtration cake can overhang neighbouring pores and influence the clog formation, there is, to our knowledge, no direct analysis of the pore cross-talk phenomenon. Yet it could have a dramatic impact on the understanding of filtration process of suspensions at macroscale, such as possible preferential locations of cake formation.  

In this work, we present observations of cross-talk when a Brownian suspension flows through a 1-D microfiltration device. 
The flow is driven by a fixed pressure difference, not a fixed flow rate, and this is a key point of our study. We measure a clogging growth rate as a function of  the number of already clogged pores and we propose a model based on a local increase of colloid concentration close to clogged pores to explain the observations. 


 \begin{center}
\begin{figure*}[t]
\includegraphics[width=1\textwidth]{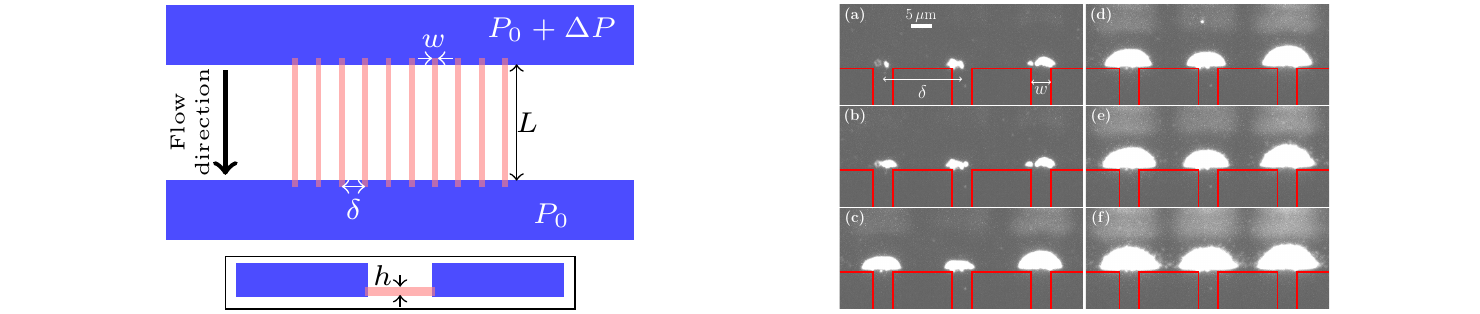}
\caption{\label{setup_photo}{Sketch of the model pores and micrograph of some clogs.} Left: top view of the chip design with zoom on nanoslits. Microchannels are represented in blue, nanoslits in red. Inset: side view of the nanoslits (not to scale). Right: image of the development of three adjacent clogs at the entrance of pores. Red lines delimit the nanoslits. For (a) to (f) the corresponding times are t=1333, 2000, 2666, 3000, 3333, 4000~s.}
\end{figure*}
\end{center}

 \textbf{Experimental method}
 
 Ten nanoslits of width $w=5$\mm, length $L=50$\mm, depth $h=830$~nm, and center-to-center spacing $\delta=20~\mu$m are etched in silicon. These nanoslits connect much larger inlet and outlet \textcolor{black}{rectangular} microchannels \textcolor{black}{acting as reservoirs (depth 23\,$\mu$m, width 100\,$\mu$m and millimetric in the third direction)}. The device is covered with a 170\mm-thick borosilicate glass plate. The design is presented in figure \ref{setup_photo}\,(left) which shows that the nanoslits connect corners of the cross-section of the microchannels.   The channels are filled with a suspension of $d_p=250$~nm diameter polystyrene particles (density 1.05 g.cm$^{-3}$). The particles are carboxylate-modified and  dispersed  in a solution  of monovalent phosphate buffered saline (PBS) diluted to an ionic strength $I=3$~mM.  The zeta potential $\zeta_p $ is measured  by laser Doppler electrophoresis,  $\zeta_p= -69$~mV ($\mathrm{pH}=7.5$).  The  volume fraction of the suspension is $\phi_0=3.8\times10^{-5}$.
A pressure difference of $\Delta P=20\pm0.02$~mbar is applied across the length of the nanoslits using a controller device. Experiments are made in dead-end and slow crossflow filtration (with velocity 0 to 9\,$\mu$m/s at 2~$\mu$m from the entrance of the pores).
 
\par

The clogging dynamics are observed using wide field fluorescence microscopy with a 40$\times$ \textcolor{black}{magnification and} 1.4 numerical aperture objective. Since the  characteristic time for clogging  is found to be about one hour we acquire images of the clog growth process at a frequency of 90 frames per minute. Figure \ref{setup_photo}\,(right) shows an example of the development of three adjacent clogs. The contour of the aggregated particle mass is detected using custom \emph{Python} scripts. From this contour analysis, we are able to determine the projected area of each clog in the field of view. 

\par

\textcolor{black}{An experimental difficulty is caused by the very low flow rate involved: the total flow rate through the ten pores before clogging is about 5\,nL.min$^{-1}$, well below the sensitivity limit of commercial flow rate sensors. Perfect watertight fitting of the chip to the pressure controller must be ensured. Also, the chips can hardly be retrieved after an experiment: a clogged chip is often discarded.}

\textbf{Experimental results}

\begin{center}
\begin{figure}[h!]
\includegraphics[width=0.4\textwidth]{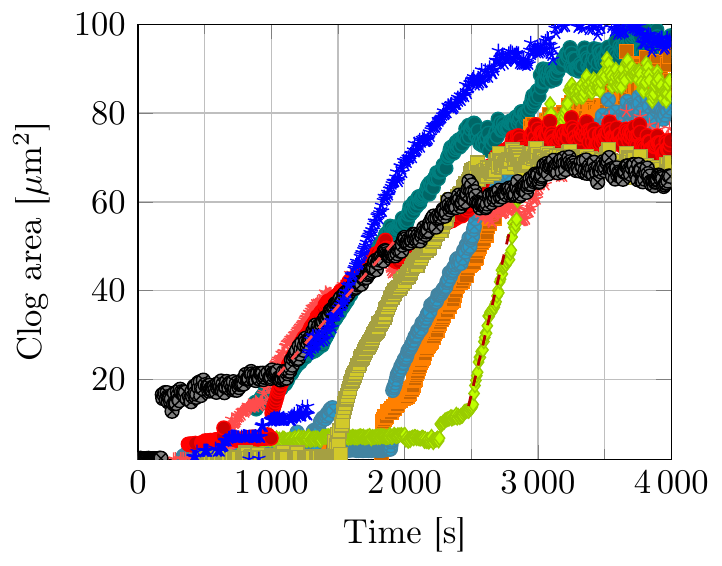}
\caption{{Time evolution of clogs area during a single acquisition.} Drawn lines highlight the zone where clog growth rate is estimated.}
\label{exemple}
\end{figure}
\end{center}

Figure \ref{exemple} shows an example of the time evolution of the area of aggregated particles at each of the pore entrances in a single microfluidic chip. 
\textcolor{black}{The curves all display the same characteristic shape. After an initial time period where the curves are quite uneven, each shows a rapid
quasi-linear growth up to a saturation level.
In the present experiment, pore clogging is mainly initiated by the capture of particle aggregates. This is not surprising as the ratio nanoslit height/particle diameter is 3.3 only. Also, the presence of small aggregates in the suspension cannot be ruled out even if care is taken to prevent the aggregate presence (by sonicating the suspension prior to its use).
The larger of these aggregates can sometimes be identified on the images, at the nanoslit entrance, once they have been captured. Aggregates  partially obstruct the pore and then initiate the slow clog growth sometimes visible at the beginning of the clogging, see e.g. the black and blue data points in figure \ref{exemple}. When the pore is fully blocked, all the particles 
 are sieved from the flow and most are captured on the aggregate (some may move laterally) so that the clog begins its fast-growth phase.  The saturation of the clog is apparently due to a balance between drag (note that the flow rate through a pore decreases when the clog grows, leading to a decrease of the drag force exerted on the particles) and the combination of double layer repulsion and Brownian diffusion, resulting in a zero particle flux surface, similar to} \textcolor{black}{the situation described by Bacchin \emph{et al.} \cite{bacchin2011}.}
\textcolor{black}{We note that this balance between transport mechanisms  is the one classically put forward to explain the existence of a stationary concentration polarization layer in filtration of colloidal suspensions \cite{song1995,bacchin2002}.
An analogous equilibrium (fluid flow-induced drag forces vs diffusiophoretic flow-induced ones) has been observed recently \cite{shin2017}. 
Note that in the case of an experiment performed with a fixed flow rate (in contrast with the present fixed pressure drop configuration), no saturation of the clog size would be observed: particles would continue to accumulate indefinitely on the clog \cite{laar2016}. }

\textcolor{black}{In the present paper, neither  a precise description  of the clogging dynamics at the pore scale nor a  quantitative  description of the saturation mechanisms are the objectives  of the study (note studies of clogging at pore scale in  conditions similar to that of the experiments  have already been performed \cite{bacchin2011,dersoir2015,robert_de_saint_vincent2016}). We rather focus on the following observation:  when a clog begins its rapid growth after other pores have reached saturation, the growth rate is larger: e.g., compare the green and black curves in figure \ref{exemple}. The goal of the paper is to describe quantitatively and to model this observation. }

\begin{center}
\begin{figure}[h!]
\includegraphics[width=0.4\textwidth]{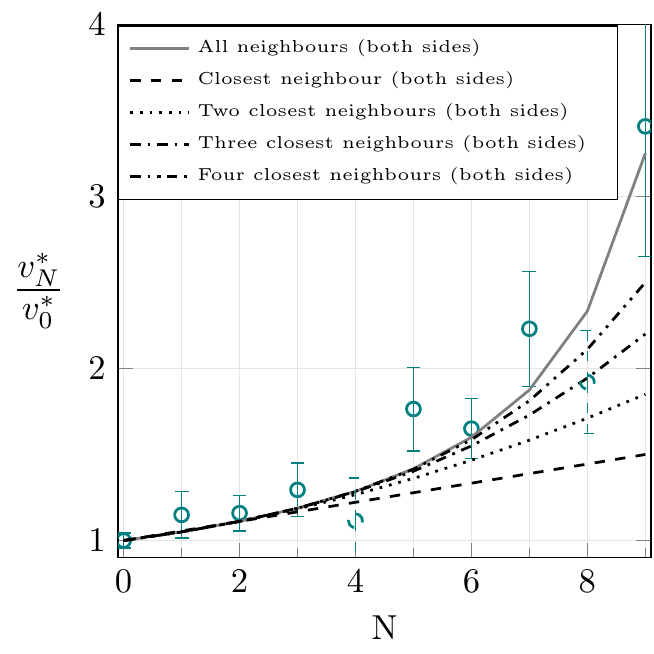}
\caption{{$v^*_N/v^*_0$ ratio versus $N$.} The error bars are related to statistical uncertainty (standard deviation over root square of the number of events); dashed points are low statistic points (less than three events). The different lines show the predictions of  eq. \ref{eq:model}\textcolor{black}{, with different ways of computing $\langle\aleph\rangle_N$, see legend for details.}}
\label{vn}
\end{figure}
\end{center}

\textcolor{black}{We quantitatively define clog growth rate (with dimensions of area/time) as the average of the derivative of the measured area with time in the zone} \textcolor{black}{starting from the beginning of the fast growth part} \textcolor{black}{where its evolution is linear, see figure \ref{exemple} (more details on the data processing are given as Supplementary materials 1.). }
We made eight acquisitions, \textcolor{black}{totaling  80 growth rate measurements.} 
We define the mean growth rate when $N$ pores are saturated (i.e.  clog size has reached saturation) as $v^*_N$. We have $N\in[ 0, N_{tot}-1]$where $N_{tot}$ is the total number of pores. \textcolor{black}{Note that during an experiment,  two or more pores may start  to clog at a very close time. Consequently, for a given experiment, all values of $N$ are not necessarily observed. As an example, if the two first pore-clogging events are simultaneous, the data for the third one will count as a $N=2$ event and such an experiment lacks a $N=1$ event.} 
\textcolor{black}{Figure \ref{vn} displays the ratio} $v^*_N/v^*_0$ as a function of $N$. Despite some large errors bars,  a clear increase of $v^*_N/v^*_0$ is observed when the number of saturated pores increases, from $v^*_N/v^*_0 =1$ for $N=0$ to $v^*_N/v^*_0 \approx 3$ for $N=9$.
\textcolor{black}{As already mentioned, explaining this increase  of $v^*_N/v^*_0$ with $N$ is the main goal of this paper and we now propose  a phenomenological model. }

\textbf{\textcolor{black}{Phenomenological} model} 

\textcolor{black}{In the present fixed pressure-drop configuration, a} saturated clog acts as a filter: \textcolor{black}{fluid is permeating through it at a flow rate $Q_s$, whereas  the flow rate through a free pore, $Q_f$, remains constant as long as clogging did not begin. Clog growth rate is proportional to this flow rate  $Q_f$ and the local concentration of particles \cite{wyss2006}.
The main idea of the present model is  that the particles driven by the permeation flow close to a clogged pore will be ``redistributed'' along the membrane surface to flow through open pores.}

\textcolor{black}{A possible mechanism to explain this redistribution is  Brownian diffusion.} The typical time scale $t^*$ between two successive pore clogging events  is about 100~s (see figure \ref{exemple}). The diffusion coefficient of the particles at ambient temperature is $D=1.7\times10^{-12}$~m$^2$.s$^{-1}$ \cite{cussler2009}. Their typical displacement  during this time interval in one direction is $\delta x=\sqrt{2Dt^*}=18~\mu$m.  \textcolor{black}{This typical length scale rises to about $60 \mathrm{\mu m}$ when considering a typical experiment duration, $\mathcal{O}(1000\,\mathrm{s})$.} Therefore, particle diffusive redistribution is expected to be a relevant mechanism in the present experiments.
\textcolor{black}{This redistribution of particles is limited: all the particles redistributed from a saturated pore will be ``sucked down'' by the first free pores (on both sides of the saturated pore). In fact, advection of a particle through an open pore is
much stronger than the diffusive transport away from it. This can be appreciated by building a P\'eclet number using the particle diameter and a typical flow velocity $U$:  $Pe= 3\,\pi\eta d_p^2 U/ k_B T$. 
A typical velocity within the nanoslit can be taken as $U_f= Q_f/(h w)$ , where $Q_f$ is computed from velocity field through a channel with rectangular cross-section with $\Delta P = 20 \, \mathrm{mbar}$ \cite{bruus2007}.
Away from the nanoslit, a typical velocity in the flow stream that will end up flowing through the nanoslit can be estimated as 
$U_f\times(h \times w) / (\delta \times 23 \, \mathrm{\mu m}$), which is $\approx 1/50$.
The P\'eclet is finally in the range 6-300, taking  the typical velocity  $U$  in the range  $U_f/50 - U_f$, showing that a particle will not be able to diffuse across a free pore, but will be captured by the flow into it.
To strengthen this point, we compute the probability, when a pore saturates, that the next one to clog is at a distance $\Delta x$ and observe a deviation from a stochastic process only for one inter-pore distance (see figure 2 in Supplementary materials).}

\begin{center}
\begin{figure}[h!]
\includegraphics[width=0.45\textwidth]{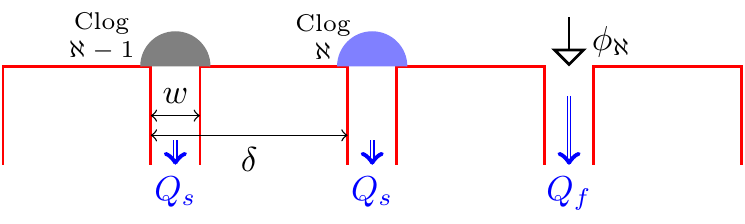}
\caption{{Sketch of a situation with two saturated pores neighbouring a free pore (therefore, $\aleph=2$).}
}
\label{schema}
\end{figure}
\end{center}

We \textcolor{black}{now} propose a stationary \textcolor{black}{phenomenological} model to estimate the influence of the redistributed particles on clog growth rate.
Assume that a free pore has $\aleph$ successive neighbouring saturated pores, including left and right directions. Figure \ref{schema} sketches the physical configuration for a $\aleph=2$ case, with the two clogs saturated on the left side of the considered free pore. 
Each saturated pore acts as a source of particles diffusing  in both directions away from the clog, toward the closest free pores. 
\textcolor{black}{Only the growth rate of these first pores will be influenced by the saturated ones.}
The effective concentration of the suspension flowing through a free pore will be simply

\begin{equation}
\phi_\aleph=\phi_0\left(1+\aleph\frac{Q_s}{2\,Q_f}\right),
\end{equation}
each neighbouring clogged pore contributing equally to $\phi_\aleph$ in the present model. 
However, the variable $\aleph$ is not easily available: it is specific to each free pore, depending on its environment which changes with time, and it depends on the configuration of free and saturated pores. Thereby, this problem has a statistical facet.
When $N$ pores out of $N_{tot}$ are saturated, the effective concentration of the suspension flowing in a given opened pore, averaged over all opened-clogged pores configurations is defined as:

\begin{equation}
\langle \phi_\aleph\rangle_N = \phi_0\left(1+\langle\aleph\rangle_N\frac{Q_s}{2\,Q_f}\right),
\end{equation}
where $\langle\aleph\rangle_N$ is the average number of neighbouring saturated pores adjacent to free pores.
 The clog growth rate is linked to the effective concentration of the suspension flowing through  the pore: $v^*_N\propto \langle \phi_\aleph\rangle_N Q_f$ \cite{wyss2006}. Finally, the ratio of clog growth rate with $N$ clogged pores to clog growth rate with no clogged pore can be written as: 
 
\begin{equation}
\frac{v_N^*}{v_0^*}=1+\langle\aleph\rangle_N\frac{Q_s}{2\,Q_f},
\label{eq:model}
\end{equation}
 with $\langle\aleph\rangle_N$ computed using a tree diagram approach (see Supplementary materials 3.).

\textcolor{black}{The flow rate $Q_f$ is determined as already explained. The flow rate $Q_s$ through a porous clog depends on its  hydraulic resistance $R_h^{clog}$, which can be estimated} using the Blake-Kozeny equation \cite{bird2002}: 
\begin{equation}
R_h^{clog}=\frac{150\,\ell\eta(1-\varepsilon)^2}{wh d_p \varepsilon^3}, 
\end{equation}
where  $\varepsilon$ represents the clog porosity, $\eta$ the fluid dynamic viscosity , and $\ell$ a typical length of the clog. To estimate $\ell$, we consider that, because they form in a corner, the clogs are roughly quarter-spheres, and assume that the radius of the clog corresponds to $\ell$. Using the projected area at saturation, we obtain $\ell\sim 7\mu$m. The porosity can be estimated from a previous study \cite{hieke2009}, which shows that for $\text{pH}=6$ (close to that in the present study), and for a small filtration cake of colloids, the porosity can reach $\varepsilon=0.83$. \textcolor{black}{This value is in good agreement with macroscopic measurements made by Brenner \cite{brenner1961}.} With these parameters, the hydraulic resistance estimate is $R_h^{clog}\approx2\times10^{17}$~kg.m$^{-4}$.s$^{-1}$. The hydraulic resistance of the rectangular pore is very similar: $R_h^{slit}=2.3\times10^{17}$~kg.m$^{-4}$.s$^{-1}$ \cite{bruus2007}. The resulting flow rate $Q_s$ through a saturated pore will be about one-half of that in a free pore\textcolor{black}{: $Q_s/Q_f \approx 0.5$}.

\textbf{Discussion and conclusion} 

The result of the model given by eq. \ref{eq:model} is plotted \textcolor{black}{as a} solid line in figure \ref{vn}.
This model does not contain free parameter: $\langle\aleph\rangle_N$ is computed numerically and the \textcolor{black}{ratio $Q_s/Q_f$ is estimated on solid grounds.  
The model is in relatively good agreement with the experimental data.} A key point, observed experimentally and well predicted by the model, is that the cross-talk between pores  becomes more and more important as $N$ increases. The clogging of a free pore will be influenced by a distant, saturated pore, if there are only saturated pores between them.
To support this point, figure \ref{vn}  \textcolor{black}{also shows  the results of the model obtained considering  only the first} \textcolor{black}{one, two, three and four} adjacent neighbour\textcolor{black}{(s)} in each direction to compute $\langle\aleph\rangle_N$. \textcolor{black}{We observe the convergence of these curves towards the full model (solid line), where the influence of all successive saturated neighbours is taken into account, } which highlights the ``long-range'' cross-talk between pores captured in the model.

To sum-up and conclude, we have directly imaged a fixed-pressure-drop filtration process of Brownian particles through  nanoslits, leading to the formation of clogs at the nanoslits entrance. The observation of  cross-talk between pores, with an increase of clog growth rate with the number of saturated pores, is the central point of this paper. \textcolor{black}{To the authors' knowledge, this is the first time such an experimental observation is reported. Such a result will certainly  impact  the current  understanding of membrane fouling dynamics. }

\textcolor{black}{To better explain the cross-talk physics, several experimental parameters should be varied. 
First, tuning the ``strength'' of Brownian diffusion will assess the role of this mechanism on the redistribution of the particles
along the membrane. Using less viscous fluids or smaller colloidal particles is a first hint. Tuning the chip geometry is a second one. For instance, an increase of the pore period $\delta$ should have a direct influence on the cross-talk: at a given Brownian diffusion magnitude, the diffusion time scale between pores may become too large to allow for cross-talk.
Another important quantity for the model is the flow rate $Q_s$ through a saturated pore. It depends on the clog saturation size and internal structure,  which both results from a balance between drag forces and  repulsive interactions between the accumulated Brownian particles. 
Tuning the internal clog structure is then an interesting perspective. This may be achieved by changing the suspension salinity. For instance, at low ionic force, the repulsive interactions between particles, and thus the ``effective volume'' occupied by each of them, will be higher. It could lead to a less dense and more permeable clog which should increase the cross-talk effect. 
Work along these directions is currently in progress, even if made tricky by the challenges that are inherent to such experiments, as discussed in the experimental method section.
Finally, the model presented in this paper} could be extended to different geometries, particularly 2-D membranes. 
\textcolor{black}{This configuration  could reduce the cross-talk effect because of the  more important number of neighbours, limited to two in our 1-D experiment (so the factor 1/2 in the  right-hand side  term of eq. 4). It will also impact the computation of 
$\langle\aleph\rangle_N$. Nevertheless,} it would be interesting to make similar measurements on 2-D membranes, such as microsieves, with different pore patterns and compare with the predictions of \textcolor{black}{the phenomenological} model.

\textbf{Acknowledgements}
 
  We acknowledge the F\'ed\'eration FERMaT and University of Toulouse (Project NEMESIS) for funding these researches. O. Liot warmly acknowledges M. Socol for his chip microfabrication and experimental help. This work was partly supported by LAAS-CNRS micro and nanotechnologies platform member of the French RENATECH network.

 
\bibliographystyle{apsrev4-1}

\bibliography{biblio}

\begin{thebibliography}{30}%
\makeatletter
\providecommand \@ifxundefined [1]{%
 \@ifx{#1\undefined}
}%
\providecommand \@ifnum [1]{%
 \ifnum #1\expandafter \@firstoftwo
 \else \expandafter \@secondoftwo
 \fi
}%
\providecommand \@ifx [1]{%
 \ifx #1\expandafter \@firstoftwo
 \else \expandafter \@secondoftwo
 \fi
}%
\providecommand \natexlab [1]{#1}%
\providecommand \enquote  [1]{``#1''}%
\providecommand \bibnamefont  [1]{#1}%
\providecommand \bibfnamefont [1]{#1}%
\providecommand \citenamefont [1]{#1}%
\providecommand \href@noop [0]{\@secondoftwo}%
\providecommand \href [0]{\begingroup \@sanitize@url \@href}%
\providecommand \@href[1]{\@@startlink{#1}\@@href}%
\providecommand \@@href[1]{\endgroup#1\@@endlink}%
\providecommand \@sanitize@url [0]{\catcode `\\12\catcode `\$12\catcode
  `\&12\catcode `\#12\catcode `\^12\catcode `\_12\catcode `\%12\relax}%
\providecommand \@@startlink[1]{}%
\providecommand \@@endlink[0]{}%
\providecommand \url  [0]{\begingroup\@sanitize@url \@url }%
\providecommand \@url [1]{\endgroup\@href {#1}{\urlprefix }}%
\providecommand \urlprefix  [0]{URL }%
\providecommand \Eprint [0]{\href }%
\providecommand \doibase [0]{http://dx.doi.org/}%
\providecommand \selectlanguage [0]{\@gobble}%
\providecommand \bibinfo  [0]{\@secondoftwo}%
\providecommand \bibfield  [0]{\@secondoftwo}%
\providecommand \translation [1]{[#1]}%
\providecommand \BibitemOpen [0]{}%
\providecommand \bibitemStop [0]{}%
\providecommand \bibitemNoStop [0]{.\EOS\space}%
\providecommand \EOS [0]{\spacefactor3000\relax}%
\providecommand \BibitemShut  [1]{\csname bibitem#1\endcsname}%
\let\auto@bib@innerbib\@empty
\bibitem [{\citenamefont {Tavakkoli}\ \emph {et~al.}(2015)\citenamefont
  {Tavakkoli}, \citenamefont {Grimes}, \citenamefont {Liu}, \citenamefont
  {Garcia}, \citenamefont {Correa}, \citenamefont {Cox},\ and\ \citenamefont
  {Vargas}}]{tavakkoli2015}%
  \BibitemOpen
  \bibfield  {author} {\bibinfo {author} {\bibfnamefont {M.}~\bibnamefont
  {Tavakkoli}}, \bibinfo {author} {\bibfnamefont {M.~R.}\ \bibnamefont
  {Grimes}}, \bibinfo {author} {\bibfnamefont {X.}~\bibnamefont {Liu}},
  \bibinfo {author} {\bibfnamefont {C.~K.}\ \bibnamefont {Garcia}}, \bibinfo
  {author} {\bibfnamefont {S.~C.}\ \bibnamefont {Correa}}, \bibinfo {author}
  {\bibfnamefont {Q.~J.}\ \bibnamefont {Cox}}, \ and\ \bibinfo {author}
  {\bibfnamefont {F.~M.}\ \bibnamefont {Vargas}},\ }\href {\doibase
  10.1021/ef502188u} {\bibfield  {journal} {\bibinfo  {journal} {Energy \&
  Fuels}\ }\textbf {\bibinfo {volume} {29}},\ \bibinfo {pages} {2890} (\bibinfo
  {year} {2015})}\BibitemShut {NoStop}%
\bibitem [{\citenamefont {Fuller}\ \emph {et~al.}(2002)\citenamefont {Fuller},
  \citenamefont {Wilhelm},\ and\ \citenamefont {Jacobson}}]{fuller2002}%
  \BibitemOpen
  \bibfield  {author} {\bibinfo {author} {\bibfnamefont {S.~B.}\ \bibnamefont
  {Fuller}}, \bibinfo {author} {\bibfnamefont {E.~J.}\ \bibnamefont {Wilhelm}},
  \ and\ \bibinfo {author} {\bibfnamefont {J.~M.}\ \bibnamefont {Jacobson}},\
  }\href {http://ieeexplore.ieee.org/abstract/document/982863/} {\bibfield
  {journal} {\bibinfo  {journal} {Journal of Microelectromechanical systems}\
  }\textbf {\bibinfo {volume} {11}},\ \bibinfo {pages} {54} (\bibinfo {year}
  {2002})}\BibitemShut {NoStop}%
\bibitem [{\citenamefont {Rothberg}(2013)}]{rothberg2013}%
  \BibitemOpen
  \bibfield  {author} {\bibinfo {author} {\bibfnamefont {M.~B.}\ \bibnamefont
  {Rothberg}},\ }\href {\doibase 10.1161/CIRCOUTCOMES.112.967778} {\bibfield
  {journal} {\bibinfo  {journal} {Circulation: Cardiovascular Quality and
  Outcomes}\ }\textbf {\bibinfo {volume} {6}},\ \bibinfo {pages} {129}
  (\bibinfo {year} {2013})}\BibitemShut {NoStop}%
\bibitem [{\citenamefont {Pang}\ \emph {et~al.}(2015)\citenamefont {Pang},
  \citenamefont {Shen}, \citenamefont {Ma}, \citenamefont {Ma}, \citenamefont
  {Zhang}, \citenamefont {Tian}, \citenamefont {Zhao}, \citenamefont {Liu},\
  and\ \citenamefont {Wang}}]{pang2015}%
  \BibitemOpen
  \bibfield  {author} {\bibinfo {author} {\bibfnamefont {L.}~\bibnamefont
  {Pang}}, \bibinfo {author} {\bibfnamefont {S.}~\bibnamefont {Shen}}, \bibinfo
  {author} {\bibfnamefont {C.}~\bibnamefont {Ma}}, \bibinfo {author}
  {\bibfnamefont {T.}~\bibnamefont {Ma}}, \bibinfo {author} {\bibfnamefont
  {R.}~\bibnamefont {Zhang}}, \bibinfo {author} {\bibfnamefont
  {C.}~\bibnamefont {Tian}}, \bibinfo {author} {\bibfnamefont {L.}~\bibnamefont
  {Zhao}}, \bibinfo {author} {\bibfnamefont {W.}~\bibnamefont {Liu}}, \ and\
  \bibinfo {author} {\bibfnamefont {J.}~\bibnamefont {Wang}},\ }\href {\doibase
  10.1039/C5AN00799B} {\bibfield  {journal} {\bibinfo  {journal} {The Analyst}\
  }\textbf {\bibinfo {volume} {140}},\ \bibinfo {pages} {7335} (\bibinfo {year}
  {2015})}\BibitemShut {NoStop}%
\bibitem [{\citenamefont {Zhang}\ \emph {et~al.}(2012)\citenamefont {Zhang},
  \citenamefont {Tang}, \citenamefont {Weisbrod},\ and\ \citenamefont
  {Guan}}]{zhang2012}%
  \BibitemOpen
  \bibfield  {author} {\bibinfo {author} {\bibfnamefont {W.}~\bibnamefont
  {Zhang}}, \bibinfo {author} {\bibfnamefont {X.}~\bibnamefont {Tang}},
  \bibinfo {author} {\bibfnamefont {N.}~\bibnamefont {Weisbrod}}, \ and\
  \bibinfo {author} {\bibfnamefont {Z.}~\bibnamefont {Guan}},\ }\href {\doibase
  10.1007/s11629-012-2443-1} {\bibfield  {journal} {\bibinfo  {journal}
  {Journal of Mountain Science}\ }\textbf {\bibinfo {volume} {9}},\ \bibinfo
  {pages} {770} (\bibinfo {year} {2012})}\BibitemShut {NoStop}%
\bibitem [{\citenamefont {Hilgers}\ and\ \citenamefont
  {Urai}(2002)}]{hilgers2002}%
  \BibitemOpen
  \bibfield  {author} {\bibinfo {author} {\bibfnamefont {C.}~\bibnamefont
  {Hilgers}}\ and\ \bibinfo {author} {\bibfnamefont {J.~L.}\ \bibnamefont
  {Urai}},\ }\href@noop {} {\bibfield  {journal} {\bibinfo  {journal} {Journal
  of Structural Geology}\ }\textbf {\bibinfo {volume} {24}},\ \bibinfo {pages}
  {1029} (\bibinfo {year} {2002})}\BibitemShut {NoStop}%
\bibitem [{\citenamefont {Dressaire}\ and\ \citenamefont
  {Sauret}(2017)}]{dressaire2017}%
  \BibitemOpen
  \bibfield  {author} {\bibinfo {author} {\bibfnamefont {E.}~\bibnamefont
  {Dressaire}}\ and\ \bibinfo {author} {\bibfnamefont {A.}~\bibnamefont
  {Sauret}},\ }\href {\doibase 10.1039/C6SM01879C} {\bibfield  {journal}
  {\bibinfo  {journal} {Soft Matter}\ }\textbf {\bibinfo {volume} {13}},\
  \bibinfo {pages} {37} (\bibinfo {year} {2017})}\BibitemShut {NoStop}%
\bibitem [{\citenamefont {Sauret}\ \emph {et~al.}(2014)\citenamefont {Sauret},
  \citenamefont {Barney}, \citenamefont {Perro}, \citenamefont {Villermaux},
  \citenamefont {Stone},\ and\ \citenamefont {Dressaire}}]{sauret2014}%
  \BibitemOpen
  \bibfield  {author} {\bibinfo {author} {\bibfnamefont {A.}~\bibnamefont
  {Sauret}}, \bibinfo {author} {\bibfnamefont {E.~C.}\ \bibnamefont {Barney}},
  \bibinfo {author} {\bibfnamefont {A.}~\bibnamefont {Perro}}, \bibinfo
  {author} {\bibfnamefont {E.}~\bibnamefont {Villermaux}}, \bibinfo {author}
  {\bibfnamefont {H.~A.}\ \bibnamefont {Stone}}, \ and\ \bibinfo {author}
  {\bibfnamefont {E.}~\bibnamefont {Dressaire}},\ }\href {\doibase
  10.1063/1.4893459} {\bibfield  {journal} {\bibinfo  {journal} {Applied
  Physics Letters}\ }\textbf {\bibinfo {volume} {105}},\ \bibinfo {pages}
  {074101} (\bibinfo {year} {2014})}\BibitemShut {NoStop}%
\bibitem [{\citenamefont {Zuriguel}\ \emph {et~al.}(2014)\citenamefont
  {Zuriguel}, \citenamefont {Parisi}, \citenamefont {Hidalgo}, \citenamefont
  {Lozano}, \citenamefont {Janda}, \citenamefont {Gago}, \citenamefont
  {Peralta}, \citenamefont {Ferrer}, \citenamefont {Pugnaloni}, \citenamefont
  {Cl{\'e}ment}, \citenamefont {Maza}, \citenamefont {Pagonabarraga},\ and\
  \citenamefont {Garcimart{\'i}n}}]{zuriguel2014}%
  \BibitemOpen
  \bibfield  {author} {\bibinfo {author} {\bibfnamefont {I.}~\bibnamefont
  {Zuriguel}}, \bibinfo {author} {\bibfnamefont {D.~R.}\ \bibnamefont
  {Parisi}}, \bibinfo {author} {\bibfnamefont {R.~C.}\ \bibnamefont {Hidalgo}},
  \bibinfo {author} {\bibfnamefont {C.}~\bibnamefont {Lozano}}, \bibinfo
  {author} {\bibfnamefont {A.}~\bibnamefont {Janda}}, \bibinfo {author}
  {\bibfnamefont {P.~A.}\ \bibnamefont {Gago}}, \bibinfo {author}
  {\bibfnamefont {J.~P.}\ \bibnamefont {Peralta}}, \bibinfo {author}
  {\bibfnamefont {L.~M.}\ \bibnamefont {Ferrer}}, \bibinfo {author}
  {\bibfnamefont {L.~A.}\ \bibnamefont {Pugnaloni}}, \bibinfo {author}
  {\bibfnamefont {E.}~\bibnamefont {Cl{\'e}ment}}, \bibinfo {author}
  {\bibfnamefont {D.}~\bibnamefont {Maza}}, \bibinfo {author} {\bibfnamefont
  {I.}~\bibnamefont {Pagonabarraga}}, \ and\ \bibinfo {author} {\bibfnamefont
  {A.}~\bibnamefont {Garcimart{\'i}n}},\ }\href {\doibase 10.1038/srep07324}
  {\bibfield  {journal} {\bibinfo  {journal} {Scientific Reports}\ }\textbf
  {\bibinfo {volume} {4}},\ \bibinfo {pages} {7324} (\bibinfo {year}
  {2014})}\BibitemShut {NoStop}%
\bibitem [{\citenamefont {Wyss}\ \emph {et~al.}(2006)\citenamefont {Wyss},
  \citenamefont {Blair}, \citenamefont {Morris}, \citenamefont {Stone},\ and\
  \citenamefont {Weitz}}]{wyss2006}%
  \BibitemOpen
  \bibfield  {author} {\bibinfo {author} {\bibfnamefont {H.~M.}\ \bibnamefont
  {Wyss}}, \bibinfo {author} {\bibfnamefont {D.~L.}\ \bibnamefont {Blair}},
  \bibinfo {author} {\bibfnamefont {J.~F.}\ \bibnamefont {Morris}}, \bibinfo
  {author} {\bibfnamefont {H.~A.}\ \bibnamefont {Stone}}, \ and\ \bibinfo
  {author} {\bibfnamefont {D.~A.}\ \bibnamefont {Weitz}},\ }\href {\doibase
  10.1103/PhysRevE.74.061402} {\bibfield  {journal} {\bibinfo  {journal}
  {Physical Review E}\ }\textbf {\bibinfo {volume} {74}} (\bibinfo {year}
  {2006}),\ 10.1103/PhysRevE.74.061402}\BibitemShut {NoStop}%
\bibitem [{\citenamefont {Dersoir}\ \emph {et~al.}(2015)\citenamefont
  {Dersoir}, \citenamefont {de~Saint~Vincent}, \citenamefont {Abkarian},\ and\
  \citenamefont {Tabuteau}}]{dersoir2015}%
  \BibitemOpen
  \bibfield  {author} {\bibinfo {author} {\bibfnamefont {B.}~\bibnamefont
  {Dersoir}}, \bibinfo {author} {\bibfnamefont {M.~R.}\ \bibnamefont
  {de~Saint~Vincent}}, \bibinfo {author} {\bibfnamefont {M.}~\bibnamefont
  {Abkarian}}, \ and\ \bibinfo {author} {\bibfnamefont {H.}~\bibnamefont
  {Tabuteau}},\ }\href {\doibase 10.1007/s10404-015-1624-y} {\bibfield
  {journal} {\bibinfo  {journal} {Microfluidics and Nanofluidics}\ }\textbf
  {\bibinfo {volume} {19}},\ \bibinfo {pages} {953} (\bibinfo {year}
  {2015})}\BibitemShut {NoStop}%
\bibitem [{\citenamefont {Bacchin}\ \emph {et~al.}(2011)\citenamefont
  {Bacchin}, \citenamefont {Marty}, \citenamefont {Duru}, \citenamefont
  {Meireles},\ and\ \citenamefont {Aimar}}]{bacchin2011}%
  \BibitemOpen
  \bibfield  {author} {\bibinfo {author} {\bibfnamefont {P.}~\bibnamefont
  {Bacchin}}, \bibinfo {author} {\bibfnamefont {A.}~\bibnamefont {Marty}},
  \bibinfo {author} {\bibfnamefont {P.}~\bibnamefont {Duru}}, \bibinfo {author}
  {\bibfnamefont {M.}~\bibnamefont {Meireles}}, \ and\ \bibinfo {author}
  {\bibfnamefont {P.}~\bibnamefont {Aimar}},\ }\href {\doibase
  10.1016/j.cis.2010.10.005} {\bibfield  {journal} {\bibinfo  {journal}
  {Advances in Colloid and Interface Science}\ }\textbf {\bibinfo {volume}
  {164}},\ \bibinfo {pages} {2} (\bibinfo {year} {2011})}\BibitemShut {NoStop}%
\bibitem [{\citenamefont {Duru}\ and\ \citenamefont {Hallez}(2015)}]{duru2015}%
  \BibitemOpen
  \bibfield  {author} {\bibinfo {author} {\bibfnamefont {P.}~\bibnamefont
  {Duru}}\ and\ \bibinfo {author} {\bibfnamefont {Y.}~\bibnamefont {Hallez}},\
  }\href {\doibase 10.1021/acs.langmuir.5b01298} {\bibfield  {journal}
  {\bibinfo  {journal} {Langmuir}\ }\textbf {\bibinfo {volume} {31}},\ \bibinfo
  {pages} {8310} (\bibinfo {year} {2015})}\BibitemShut {NoStop}%
\bibitem [{\citenamefont {Robert~de Saint~Vincent}\ \emph
  {et~al.}(2016)\citenamefont {Robert~de Saint~Vincent}, \citenamefont
  {Abkarian},\ and\ \citenamefont {Tabuteau}}]{robert_de_saint_vincent2016}%
  \BibitemOpen
  \bibfield  {author} {\bibinfo {author} {\bibfnamefont {M.}~\bibnamefont
  {Robert~de Saint~Vincent}}, \bibinfo {author} {\bibfnamefont
  {M.}~\bibnamefont {Abkarian}}, \ and\ \bibinfo {author} {\bibfnamefont
  {H.}~\bibnamefont {Tabuteau}},\ }\href {\doibase 10.1039/C5SM01952D}
  {\bibfield  {journal} {\bibinfo  {journal} {Soft Matter}\ }\textbf {\bibinfo
  {volume} {12}},\ \bibinfo {pages} {1041} (\bibinfo {year}
  {2016})}\BibitemShut {NoStop}%
\bibitem [{\citenamefont {Dersoir}\ \emph {et~al.}(2017)\citenamefont
  {Dersoir}, \citenamefont {Schofield},\ and\ \citenamefont
  {Tabuteau}}]{dersoir2017}%
  \BibitemOpen
  \bibfield  {author} {\bibinfo {author} {\bibfnamefont {B.}~\bibnamefont
  {Dersoir}}, \bibinfo {author} {\bibfnamefont {A.~B.}\ \bibnamefont
  {Schofield}}, \ and\ \bibinfo {author} {\bibfnamefont {H.}~\bibnamefont
  {Tabuteau}},\ }\href {\doibase 10.1039/C6SM02605B} {\bibfield  {journal}
  {\bibinfo  {journal} {Soft Matter}\ }\textbf {\bibinfo {volume} {13}},\
  \bibinfo {pages} {2054} (\bibinfo {year} {2017})}\BibitemShut {NoStop}%
\bibitem [{\citenamefont {Cejas}\ \emph {et~al.}(2017)\citenamefont {Cejas},
  \citenamefont {Monti}, \citenamefont {Truchet}, \citenamefont {Burnouf},\
  and\ \citenamefont {Tabeling}}]{cejas2017}%
  \BibitemOpen
  \bibfield  {author} {\bibinfo {author} {\bibfnamefont {C.~M.}\ \bibnamefont
  {Cejas}}, \bibinfo {author} {\bibfnamefont {F.}~\bibnamefont {Monti}},
  \bibinfo {author} {\bibfnamefont {M.}~\bibnamefont {Truchet}}, \bibinfo
  {author} {\bibfnamefont {J.-P.}\ \bibnamefont {Burnouf}}, \ and\ \bibinfo
  {author} {\bibfnamefont {P.}~\bibnamefont {Tabeling}},\ }\href {\doibase
  10.1021/acs.langmuir.7b01394} {\bibfield  {journal} {\bibinfo  {journal}
  {Langmuir}\ }\textbf {\bibinfo {volume} {33}},\ \bibinfo {pages} {6471}
  (\bibinfo {year} {2017})}\BibitemShut {NoStop}%
\bibitem [{\citenamefont {Kim}\ \emph {et~al.}(2017)\citenamefont {Kim},
  \citenamefont {Ahn},\ and\ \citenamefont {Lee}}]{kim2017}%
  \BibitemOpen
  \bibfield  {author} {\bibinfo {author} {\bibfnamefont {Y.}~\bibnamefont
  {Kim}}, \bibinfo {author} {\bibfnamefont {K.~H.}\ \bibnamefont {Ahn}}, \ and\
  \bibinfo {author} {\bibfnamefont {S.~J.}\ \bibnamefont {Lee}},\ }\href
  {\doibase 10.1016/j.memsci.2017.04.010} {\bibfield  {journal} {\bibinfo
  {journal} {Journal of Membrane Science}\ }\textbf {\bibinfo {volume} {534}},\
  \bibinfo {pages} {25} (\bibinfo {year} {2017})}\BibitemShut {NoStop}%
\bibitem [{\citenamefont {Laar}\ \emph {et~al.}(2016)\citenamefont {Laar},
  \citenamefont {Klooster}, \citenamefont {Schro{\"e}n},\ and\ \citenamefont
  {Sprakel}}]{laar2016}%
  \BibitemOpen
  \bibfield  {author} {\bibinfo {author} {\bibfnamefont {T.~v.~d.}\
  \bibnamefont {Laar}}, \bibinfo {author} {\bibfnamefont {S.~t.}\ \bibnamefont
  {Klooster}}, \bibinfo {author} {\bibfnamefont {K.}~\bibnamefont
  {Schro{\"e}n}}, \ and\ \bibinfo {author} {\bibfnamefont {J.}~\bibnamefont
  {Sprakel}},\ }\href {\doibase 10.1038/srep28450} {\bibfield  {journal}
  {\bibinfo  {journal} {Scientific Reports}\ }\textbf {\bibinfo {volume} {6}}
  (\bibinfo {year} {2016}),\ 10.1038/srep28450}\BibitemShut {NoStop}%
\bibitem [{\citenamefont {Sendekie}\ and\ \citenamefont
  {Bacchin}(2016)}]{sendekie2016}%
  \BibitemOpen
  \bibfield  {author} {\bibinfo {author} {\bibfnamefont {Z.~B.}\ \bibnamefont
  {Sendekie}}\ and\ \bibinfo {author} {\bibfnamefont {P.}~\bibnamefont
  {Bacchin}},\ }\href {\doibase 10.1021/acs.langmuir.5b04218} {\bibfield
  {journal} {\bibinfo  {journal} {Langmuir}\ }\textbf {\bibinfo {volume}
  {32}},\ \bibinfo {pages} {1478} (\bibinfo {year} {2016})}\BibitemShut
  {NoStop}%
\bibitem [{\citenamefont {Brenner}(1961)}]{brenner1961}%
  \BibitemOpen
  \bibfield  {author} {\bibinfo {author} {\bibfnamefont {H.}~\bibnamefont
  {Brenner}},\ }\href@noop {} {\bibfield  {journal} {\bibinfo  {journal} {AIChE
  Journal}\ }\textbf {\bibinfo {volume} {7}},\ \bibinfo {pages} {666} (\bibinfo
  {year} {1961})}\BibitemShut {NoStop}%
\bibitem [{\citenamefont {Ghidaglia}\ \emph {et~al.}(1996)\citenamefont
  {Ghidaglia}, \citenamefont {de~Arcangelis}, \citenamefont {Hinch},\ and\
  \citenamefont {Guazzelli}}]{ghidaglia1996}%
  \BibitemOpen
  \bibfield  {author} {\bibinfo {author} {\bibfnamefont {C.}~\bibnamefont
  {Ghidaglia}}, \bibinfo {author} {\bibfnamefont {L.}~\bibnamefont
  {de~Arcangelis}}, \bibinfo {author} {\bibfnamefont {J.}~\bibnamefont
  {Hinch}}, \ and\ \bibinfo {author} {\bibfnamefont {{\'E}.}~\bibnamefont
  {Guazzelli}},\ }\href
  {https://journals.aps.org/pre/abstract/10.1103/PhysRevE.53.R3028} {\bibfield
  {journal} {\bibinfo  {journal} {Physical Review E}\ }\textbf {\bibinfo
  {volume} {53}},\ \bibinfo {pages} {R3028} (\bibinfo {year}
  {1996})}\BibitemShut {NoStop}%
\bibitem [{\citenamefont {Narayan}\ \emph {et~al.}(1997)\citenamefont
  {Narayan}, \citenamefont {Coury}, \citenamefont {Masliyah},\ and\
  \citenamefont {Gray}}]{narayan1997}%
  \BibitemOpen
  \bibfield  {author} {\bibinfo {author} {\bibfnamefont {R.}~\bibnamefont
  {Narayan}}, \bibinfo {author} {\bibfnamefont {J.~R.}\ \bibnamefont {Coury}},
  \bibinfo {author} {\bibfnamefont {J.~H.}\ \bibnamefont {Masliyah}}, \ and\
  \bibinfo {author} {\bibfnamefont {M.~R.}\ \bibnamefont {Gray}},\ }\href
  {http://pubs.acs.org/doi/abs/10.1021/ie970101e} {\bibfield  {journal}
  {\bibinfo  {journal} {Industrial \& engineering chemistry research}\ }\textbf
  {\bibinfo {volume} {36}},\ \bibinfo {pages} {4620} (\bibinfo {year}
  {1997})}\BibitemShut {NoStop}%
\bibitem [{\citenamefont {Hong}\ \emph {et~al.}(1997)\citenamefont {Hong},
  \citenamefont {Faibish},\ and\ \citenamefont {Elimielech}}]{hong1997}%
  \BibitemOpen
  \bibfield  {author} {\bibinfo {author} {\bibfnamefont {S.}~\bibnamefont
  {Hong}}, \bibinfo {author} {\bibfnamefont {R.~S.}\ \bibnamefont {Faibish}}, \
  and\ \bibinfo {author} {\bibfnamefont {M.}~\bibnamefont {Elimielech}},\
  }\href@noop {} {\bibfield  {journal} {\bibinfo  {journal} {Journal of Colloid
  ans Interface Science}\ }\textbf {\bibinfo {volume} {196}},\ \bibinfo {pages}
  {267} (\bibinfo {year} {1997})}\BibitemShut {NoStop}%
\bibitem [{\citenamefont {Hieke}\ \emph {et~al.}(2009)\citenamefont {Hieke},
  \citenamefont {Ruland}, \citenamefont {Anlauf},\ and\ \citenamefont
  {Nirschl}}]{hieke2009}%
  \BibitemOpen
  \bibfield  {author} {\bibinfo {author} {\bibfnamefont {M.}~\bibnamefont
  {Hieke}}, \bibinfo {author} {\bibfnamefont {J.}~\bibnamefont {Ruland}},
  \bibinfo {author} {\bibfnamefont {H.}~\bibnamefont {Anlauf}}, \ and\ \bibinfo
  {author} {\bibfnamefont {H.}~\bibnamefont {Nirschl}},\ }\href {\doibase
  10.1002/ceat.200800609} {\bibfield  {journal} {\bibinfo  {journal} {Chemical
  Engineering \& Technology}\ }\textbf {\bibinfo {volume} {32}},\ \bibinfo
  {pages} {1095} (\bibinfo {year} {2009})}\BibitemShut {NoStop}%
\bibitem [{\citenamefont {Song}\ and\ \citenamefont
  {Elimelech}(1995)}]{song1995}%
  \BibitemOpen
  \bibfield  {author} {\bibinfo {author} {\bibfnamefont {L.}~\bibnamefont
  {Song}}\ and\ \bibinfo {author} {\bibfnamefont {M.}~\bibnamefont
  {Elimelech}},\ }\href@noop {} {\bibfield  {journal} {\bibinfo  {journal}
  {Journal of the Chemical Society, Faraday Transactions}\ }\textbf {\bibinfo
  {volume} {91}},\ \bibinfo {pages} {3389} (\bibinfo {year}
  {1995})}\BibitemShut {NoStop}%
\bibitem [{\citenamefont {Bacchin}\ \emph {et~al.}(2002)\citenamefont
  {Bacchin}, \citenamefont {Si-Hassen}, \citenamefont {Starov}, \citenamefont
  {Clifton},\ and\ \citenamefont {Aimar}}]{bacchin2002}%
  \BibitemOpen
  \bibfield  {author} {\bibinfo {author} {\bibfnamefont {P.}~\bibnamefont
  {Bacchin}}, \bibinfo {author} {\bibfnamefont {D.}~\bibnamefont {Si-Hassen}},
  \bibinfo {author} {\bibfnamefont {V.}~\bibnamefont {Starov}}, \bibinfo
  {author} {\bibfnamefont {M.~J.}\ \bibnamefont {Clifton}}, \ and\ \bibinfo
  {author} {\bibfnamefont {P.}~\bibnamefont {Aimar}},\ }\href {\doibase
  10.1016/S0009-2509(01)00316-5} {\bibfield  {journal} {\bibinfo  {journal}
  {Chemical Engineering Science}\ }\textbf {\bibinfo {volume} {57}},\ \bibinfo
  {pages} {77} (\bibinfo {year} {2002})}\BibitemShut {NoStop}%
\bibitem [{\citenamefont {Shin}\ \emph {et~al.}(2017)\citenamefont {Shin},
  \citenamefont {Ault}, \citenamefont {Warren},\ and\ \citenamefont
  {Stone}}]{shin2017}%
  \BibitemOpen
  \bibfield  {author} {\bibinfo {author} {\bibfnamefont {S.}~\bibnamefont
  {Shin}}, \bibinfo {author} {\bibfnamefont {J.~T.}\ \bibnamefont {Ault}},
  \bibinfo {author} {\bibfnamefont {P.~B.}\ \bibnamefont {Warren}}, \ and\
  \bibinfo {author} {\bibfnamefont {H.~A.}\ \bibnamefont {Stone}},\ }\href
  {\doibase 10.1103/PhysRevX.7.041038} {\bibfield  {journal} {\bibinfo
  {journal} {Physical Review X}\ }\textbf {\bibinfo {volume} {7}} (\bibinfo
  {year} {2017}),\ 10.1103/PhysRevX.7.041038}\BibitemShut {NoStop}%
\bibitem [{\citenamefont {Cussler}(2009)}]{cussler2009}%
  \BibitemOpen
  \bibfield  {author} {\bibinfo {author} {\bibfnamefont {E.~L.}\ \bibnamefont
  {Cussler}},\ }\href@noop {} {\emph {\bibinfo {title} {Diffusion: {Mass}
  {Transfer} in {Fluid} {Systems}}}}\ (\bibinfo  {publisher} {Cambridge
  University Press},\ \bibinfo {year} {2009})\ \bibinfo {note}
  {google-Books-ID: dq6LdJyN8ScC}\BibitemShut {NoStop}%
\bibitem [{\citenamefont {Bruus}(2007)}]{bruus2007}%
  \BibitemOpen
  \bibfield  {author} {\bibinfo {author} {\bibfnamefont {H.}~\bibnamefont
  {Bruus}},\ }\href
  {http://web-files.ait.dtu.dk/bruus/TMF/publications/books/Bruus_TMFbook_Sample_Chapter.pdf}
  {\emph {\bibinfo {title} {Theoretical microfluidics}}}\ (\bibinfo
  {publisher} {Oxford university press Oxford},\ \bibinfo {year}
  {2007})\BibitemShut {NoStop}%
\bibitem [{\citenamefont {Bird}(2002)}]{bird2002}%
  \BibitemOpen
  \bibfield  {author} {\bibinfo {author} {\bibfnamefont {R.~B.}\ \bibnamefont
  {Bird}},\ }\href {\doibase 10.1115/1.1424298} {\bibfield  {journal} {\bibinfo
   {journal} {Applied Mechanics Reviews}\ }\textbf {\bibinfo {volume} {55}},\
  \bibinfo {pages} {R1} (\bibinfo {year} {2002})}\BibitemShut {NoStop}%
\end{thebibliography}%
 

\end{document}